\documentclass[fleqn,twoside]{article}
\usepackage{espcrc2}

\usepackage{graphicx}
\usepackage[figuresright]{rotating}

\newcommand{\AmS}{{\protect\the\textfont2
  A\kern-.1667em\lower.5ex\hbox{M}\kern-.125emS}}

\title{A new light boson from MAGIC observations?}

\author{Marco Roncadelli\address{INFN, Sezione di Pavia, via A. Bassi 6, I -- 27100 Pavia, Italy} 
        Alessandro De Angelis\address{Dipartimento di Fisica, Universit\`a di Udine, Via delle Scienze 208, I -- 33100 Udine, and INAF and INFN, 
        Sezioni di Trieste, Italy} 
        and
        Oriana Mansutti\address{Dipartimento di Fisica, Universit\`a di Udine, Via delle Scienze 208, I -- 33100 Udine, and INFN, Sezione di Trieste, Italy}}
       
\begin{document}

\begin{abstract}
Recent detection of blazar 3C279 by MAGIC has confirmed previous indications by H.E.S.S. that the Universe is more transparent to very-high-energy gamma rays than currently thought. This circumstance can be reconciled with observations of nearby blazars provided that photon oscillations into a very light Axion-Like Particle occur in extragalactic magnetic fields. The emerging ``DARMA scenario'' can be tested in the near future by the satellite-borne {\it Fermi} LAT detector as well as by the ground-based Imaging Atmospheric Cherenkov Telescopes H.E.S.S., MAGIC, CANGAROO III, VERITAS and by the Extensive Air Shower arrays ARGO-YBJ and MILAGRO.
\vspace{1pc}
\end{abstract}

\maketitle

\section{MOTIVATION}

Imaging Atmospheric Cherenkov Telescopes (IACTs) are providing us with an impressive amount of information about the Universe in the energy interval $100 \, {\rm GeV} - 100 \,{\rm TeV}$. Observations carried out by these IACTs concern gamma-ray sources over an extremely wide interval of distances, ranging from the parsec scale for Galactic objects up to the Gigaparsec scale for the fartest detected blazar 3C279. This circumstance allows not only to infer the intrinsec properties of the sources, but also to probe the nature of photon propagation throughout cosmic distances. 

The latter fact is of paramount importance for very-high-energy (VHE) gamma-ray astrophysics, since the horizon of the observable Universe rapidly shrinks above $100 \, {\rm GeV}$ as the energy further increases. This is due to the fact that photons from distant sources scatter off background photons permeating the Universe, thereby disappearing into electron-positron pairs~\cite{stecker1971}. It turns out that the corresponding cross section $\sigma (\gamma \gamma \to e^+ e^-)$ peaks where the VHE photon energy $E$ and the background photon energy $\epsilon$ are related by $\epsilon \simeq (500 \, {\rm GeV}/E) \, {\rm eV}$. As far as observations performed by IACTs  are concerned, the cosmic opacity is dominated by the interaction with ultraviolet/optical/infrared diffuse background photons~\footnote{Frequency band $1.2 \cdot 10^{3} \, {\rm GHz} - 1.2 \cdot 10^{6} \, {\rm GHz}$, corresponding to the wavelength range $0.25 \, \mu {\rm m} - 250 \, \mu {\rm m}$.}, usually called Extragalactic Background Light (EBL), which is produced by galaxies during the whole history of the Universe. Owing to such an absorption process, photon propagation is controlled by the optical depth ${\tau}(E,D)$, with $D$ denoting the source distance. Hence, the observed photon flux $\Phi_{\rm obs}(E,D)$ is related to the emitted one $\Phi_{\rm em}(E)$ by 
\begin{equation}
\label{a0}
\Phi_{\rm obs}(E,D) = e^{- \tau(E,D)} \, \Phi_{\rm em}(E)~. 
\end{equation}
Neglecting evolutionary effects on the EBL spectral energy distribution for simplicity, the optical depth reads ${\tau}(E,D) \simeq D/{\lambda}_{\gamma}(E)$, where ${\lambda}_{\gamma}(E)$ is the photon mean free path for $\gamma \gamma \to e^+ e^-$ referring to the present cosmic epoch. As a consequence, 
Eq.~(\ref{a0}) simplifies as
\begin{equation}
\label{a1}
\Phi_{\rm obs}(E,D) \simeq e^{- D/{\lambda}_{\gamma}(E)} \ \Phi_{\rm em}(E)~.
\end{equation}
The function ${\lambda}_{\gamma}(E)$ decreases like a power law from the Hubble radius $4.3 \, {\rm Gpc}$ around $100 \, {\rm GeV}$ to $1 \, {\rm Mpc}$ around $100 \, {\rm TeV}$~\cite{CoppiAharonian}. Now, Eq.~(\ref{a1}) entails that the observed flux is {\it exponentially} suppressed both at high energy and at large distances, so that sufficiently far-away sources become hardly visible in the VHE range and their observed spectrum should anyway be {\it much steeper} than the emitted one.

Yet, observations carried out by IACTs have failed to detect such a behaviour. A first indication in this respect was reported by the H.E.S.S. collaboration in connection with the discovery of the two blazars H2356-309 ($z = 0.165$) and 1ES1101-232 ($z = 0.186$) at $E \sim 1 \, {\rm TeV}$~\cite{aharonian:nature06}. Stronger evidence comes from the observation of the blazar 3C279 ($z = 0.538$) at $E \sim 0.5 \, {\rm TeV}$ by the MAGIC collaboration~\cite{3c}. In particular, the signal from 3C279 collected by MAGIC in the region $E<220$ GeV has more or less the same statistical significance as the one in the range 220 GeV $< E <$ 600 GeV ($6.1 \sigma$ in the former case, $5.1 \sigma$ in the latter). 

Turning the argument around and assuming {\it standard} photon propagation as described above, the observed spectrum $\Phi_{\rm obs}(E,D)$ can only be reproduced by an emission spectrum $\Phi_{\rm em}(E)$ {\it much harder} than for any other blazar observed so far. 

A way out of this difficulty relies upon a modification of the emission spectrum $\Phi_{\rm em}(E)$. A possibility involves the presence of strong relativistic shocks, which can substantially harden $\Phi_{\rm em}(E)$~\cite{Stecker2007+2008}. A different option invokes photon absorption inside the blazar, which has been shown to produce again an emission spectrum $\Phi_{\rm em}(E)$ considerably harder than previously thought~\cite{Aharonian2008}. While successful at increasing the fraction of VHE emitted photons, these attempts fail to explain why {\it only} for the most distant blazars do these mechanisms become important.

A very different solution was recently proposed by the present authors and is usually referred to as the ``DARMA scenario''~\cite{drm}. Its characteristic feature is the presence of Axion-Like Particles (ALPs) (more about this, later) and rests upon the mechanism of photon-ALP oscillation in cosmic magnetic fields, whose existence has definitely  been proved by AUGER observations~\cite{auger}. Once ALPs are produced close enough to the source, they travel {\it unimpeded} throughout the Universe -- since they do not undergo EBL absorption -- and can convert back to photons before reaching the Earth. As a consequence, the {\it effective} photon mean free path ${\lambda}_{\gamma , {\rm eff}} (E,D)$ gets {\it increased} so that the observed photons cross a distance in excess of ${\lambda}_{\gamma}(E)$. Moreover, it has been shown that the DARMA scenario works for an ALP lighter than about $10^{- 10} \, {\rm eV}$~\footnote{Somewhat similar ideas are discussed in~\cite{sim}.}.

A deeper insight into the DARMA mechanism can be achieved by introducing the probability $P_{\gamma \to \gamma}(E,D)$ that a photon remains a photon after propagation over a distance $D$, so that we have
\begin{equation}
\label{a1bisa}
\Phi_{\rm obs}(E,D) = P_{\gamma \to \gamma}(E,D) \ \Phi_{\rm em}(E)~.
\end{equation}
When only photon absorption is operative, Eq.~(\ref{a1}) can similarly be rewritten as
\begin{equation}
\label{a1bisaQ}
\Phi_{\rm obs}(E,D) = P_{\gamma \to \gamma}^{(0)}(E,D) \ \Phi_{\rm em}(E)~,
\end{equation}
with
\begin{equation}
\label{a1trisaW}
P_{\gamma \to \gamma}^{(0)}(E,D) \simeq e^{- D/{\lambda}_{\gamma}(E)}~. 
\end{equation}
In the presence of photon-ALP oscillations, Eq.~(\ref{a1trisaW}) gets replaced by
\begin{equation}
\label{a1trisb}
P_{\gamma \to \gamma}(E,D) \simeq e^{- D/{\lambda}_{\gamma}(E)} \, X(E,D) 
\end{equation}
and the above discussion entails $X(E,D) > 1$. Moreover, Eq.~(\ref{a1}) presently becomes
\begin{equation}
\label{a1bis}
\Phi_{\rm obs}(E,D) \simeq e^{- D/{\lambda}_{\gamma , {\rm eff}}(E,D)} \ \Phi_{\rm em}(E)~,
\end{equation}
with
\begin{equation}
\label{a1tris}
{\lambda}_{\gamma , {\rm eff}}(E,D) = - \frac{D}{{\rm ln} \, P_{\gamma \to \gamma}(E,D)}~,
\end{equation}
so as to guarantee consistency with Eq.~(\ref{a1bisa}). Next, by inserting Eq.~(\ref{a1trisb}) into Eq.~(\ref{a1tris}) we get
\begin{equation}
\label{a1tric}
\frac{{\lambda}_{\gamma , {\rm eff}}(E,D)}{{\lambda}_{\gamma}(E)} \simeq \frac{D}{D - {\lambda}_{\gamma}(E) \,  {\rm ln} \, X(E,D)} 
\end{equation}
and since $X(E,D) > 1$ we find ${\lambda}_{\gamma , {\rm eff}}(E,D) > {\lambda}_{\gamma}(E)$, which is just a formal restatement of our previous conclusion. Still, Eq.~(\ref{a1bis}) possesses the advantage to explicitly show that even a {\it small} increase of ${\lambda}_{\gamma , {\rm eff}}(E,D)$ gives rise to a {\it large} enhancement of the observed flux $\Phi_{\rm obs} (E,D)$. As we shall see, the DARMA mechanism makes ${\lambda}_{\gamma , {\rm eff}}(E,D)$ shallower than ${\lambda}_{\gamma}(E)$, although it remains a decreasing function of $E$. So, the resulting observed spectrum is {\it much harder} than the one predicted by Eq.~(\ref{a1}), thereby ensuring agreement with observations even by adopting for far-away sources the {\it same} emission spectrum characteristic of nearby ones.

Our aim is to review the main features of the DARMA scenario as well as its application to blazar 3C279.

\section{DARMA SCENARIO}

Both phenomenological and conceptual arguments lead to view the Standard Model of particle physics as the low-energy manifestation of some more fundamental and richer theory of all elementary-particle interactions including gravity. Therefore, the lagrangian of the Standard Model is expected to be modified by small terms describing interactions among known and new particles. Many extensions of the Standard Model which have attracted considerable interest over the last few years indeed predict the existence of ALPs. They are spin-zero light bosons defined by the low-energy effective lagrangian
\begin{equation}
\label{a1a}
{\cal L}_{\rm ALP} \ = \ 
\frac{1}{2} \, \partial^{\mu} \, a \, \partial_{\mu} \, a - \frac{m^2}{2} 
\, a^2 - \frac{1}{4 M} \, F^{\mu \nu} \, \tilde F_{\mu \nu} \, a~,
\end{equation}
where $F^{\mu \nu}$ is the electromagnetic field strength, $\tilde F_{\mu \nu}$ is its dual, $a$ denotes the ALP field whereas $m$ stands for the ALP mass~\footnote{As usual, natural Lorentz-Heaviside units with $\hbar=c=1$ are employed throughout.}. According to the above view, it is assumed $M \gg G_F^{- 1/2} \simeq 250 \, {\rm GeV}$. On the other hand, it is supposed that $m \ll G_F^{- 1/2} \simeq 250 \, {\rm GeV}$. The standard Axion~\cite{Raffelt1990} is the most well known example of ALP. As far as {\it generic} ALPs are concerned, the parameters $M$ and $m$ are to be regarded as {\it independent}.

So, what really characterizes ALPs is the trilinear $\gamma$-$\gamma$-$a$ vertex described by the last term in ${\cal L}_{\rm ALP}$, whereby one ALP couples to two photons. Owing to this vertex, ALPs can be emitted by astronomical objects of various kinds, and the present situation can be summarized as follows. The negative result of the CAST experiment designed to detect ALPs emitted by the Sun yields the bound $M > 0.86 \cdot 10^{10} \, {\rm GeV}$ for 
$m < 0.02 \, {\rm eV}$~\cite{Zioutas2005}. Moreover, theoretical considerations concerning star cooling via ALP emission provide the generic bound $M > 10^{10} \, {\rm GeV}$, which for $m < 10^{- 10} \, {\rm eV}$ gets replaced by the stronger one $M >  10^{11} \, {\rm GeV}$  even if with a large uncertainty~\cite{Raffelt1990}. The same $\gamma$-$\gamma$-$a$ vertex produces an off-diagonal element in the mass matrix for the photon-ALP system in the presence of an external magnetic field ${\bf B}$. Therefore, the interaction eigenstates differ from the propagation eigenstates and photon-ALP oscillations show up~\cite{Sikivie1984}.

We imagine that a sizeable fraction of photons emitted by a blazar convert into ALPs because of cosmic magnetic fields (CMFs), whose existence has been demonstrated very recently by AUGER observations~\cite{auger}. These ALPs propagate unaffected by the EBL and we suppose that a substantial fraction of them back converts into photons before reaching the Earth  ALPs. Owing to the notorious lack of information about the morphology of CMFs, one usually supposes that they have a domain-like structure~\cite{Kronberg}. That is, ${\bf B}$ ought to be constant over a domain of size $L_{\rm dom}$ equal to its coherence length, with ${\bf B}$ randomly changing its direction from one domain to another but keeping approximately the same strength. As explained elsewhere~\cite{dpr}, it looks plausible to assume the coherence length in the range $1 \, {\rm Mpc} - 10 \, {\rm Mpc}$. Correspondingly, the inferred strength lies in the range $0.3 \, {\rm nG} - 1.0 \, {\rm nG}$~\cite{dpr}.

\section{PREDICTED ENERGY SPECTRUM}

Our ultimate goal consists in the evaluation of the probability $P_{\gamma \to \gamma}(E,D)$ when allowance is made for photon-ALP oscillations as well as for photon absorption from the EBL. We proceed as follows. We first solve exactly the beam propagation equation arising from ${\cal L}_{\rm ALP}$ over a single domain, assuming that the EBL is described by the ``best-fit model'' of Kneiske {\it et al.}~\cite{kneiske}. Starting with an unpolarized photon beam, we next propagate it by iterating the single-domain solution as many times as the number of domains crossed by the beam, taking each time a {\it random} value for the angle between ${\bf B}$ and a fixed overall fiducial direction. We repeat such a procedure $10^.000$ times and finally we average over all these realizations of the propagation process. 

We find that about 13\% of the photons arrive to the Earth for $E = 500 \, {\rm GeV}$, representing an enhancement by a factor of about 20 with respect to the expected flux without DARMA mechanism (the comparison is made with the above ``best-fit model''). The same calculation gives a fraction of 76\% for $E = 100 \, {\rm GeV}$ (to be compared to 67\% without DARMA mechanism) and a fraction of 3.4\% for $E =  1 \, {\rm TeV}$ (to be compared to 0.0045\% without DARMA mechanism). The resulting spectrum is exhibited in Fig.~1. The solid line represents the prediction of the DARMA scenario for  \mbox{$B \simeq 1 \, {\rm nG}$} and \mbox{$L_{\rm dom} \simeq 1 \, {\rm Mpc}$} and the gray band is the envelope of the results obtained by independently varying ${\bf B}$ and $L_{\rm dom}$ within a factor of 10 about such values. These conclusions hold for $m \ll 10^{-10} \, {\rm eV}$ and we have taken for definiteness $M \simeq 4 \cdot 10^{11} \, {\rm GeV}$ but we have cheked that practically nothing changes for $10^{11} \, {\rm GeV} < M < 10^{13} \, {\rm GeV}$.

\begin{figure}[htb]
\begin{center}
\includegraphics[width=8cm]{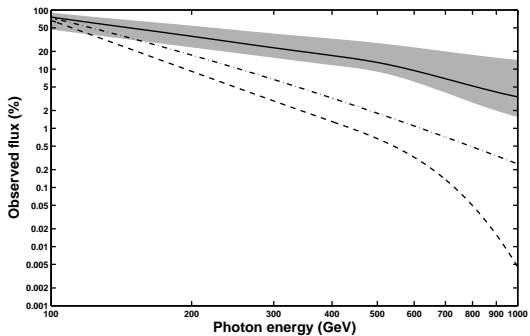}
\caption{\label{fig1}{The two lowest lines give the fraction of photons surviving from 3C279 without the DARMA mechanism within the ``best-fit model'' of EBL (dashed line) and for the minimum EBL density compatible with cosmology (dashed-dotted line). The solid line represents the prediction of the DARMA mechanism as explained in the text.}}
\end{center}
\end{figure}

Our prediction can be tested in the near future by the satellite-borne {\it Fermi} LAT detector as well as by the ground-based IACTs H.E.S.S., MAGIC, CANGAROO III, VERITAS and by the Extensive Air Shower arrays ARGO-YBJ and MILAGRO.

\end{document}